\documentclass[10pt,letterpaper,twocolumn]{article}

\usepackage{ol2}
\usepackage[draft]{hyperref}
\usepackage{amsmath}

\begin{document}

\twocolumn[

\title{Surface Acoustic Wave Frequency Comb}

\author{A. A. Savchenkov, A. B. Matsko, V. S. Ilchenko, D. Seidel, and L. Maleki}

\address{OEwaves Inc., 2555 East Colorado Boulevard, Pasadena, CA 91107 \\
$^*$Corresponding author: andrey.matsko@oewaves.com }

\begin{abstract}
We report on realization of an efficient triply-resonant coupling between two long lived optical modes and a high frequency surface acoustic wave (SAW) mode of the same monolithic crystalline whispering gallery mode resonator. The coupling results in an opto-mechanical oscillation and generation of a monochromatic SAW. A strong nonlinear interaction of this mechanical mode with other equidistant SAW modes leads to mechanical hyper-parametric oscillation and generation of a SAW pulse train and associated frequency comb in the resonator. We visualized the comb observing the modulation of the modulated light escaping the resonator.
\end{abstract}


 ]

Cavity opto-mechanics has enjoyed intense interest in the research community engaged with investigation of quantum effects in macroscopically large objects, ultra-sensitive measurement of dynamic back action, and cooling thermally excited random motion of mechanical modes of the cavity \cite{kippenberg08s}.  The study of the interaction of the optical and mechanical cavity modes has been basically limited to the mechanical oscillations that are localized in the bulk of the cavity. It was pointed out that triply-resonant opto-mechanical interaction with surface acoustic waves (SAWs) is also possible in monolithic whispering gallery mode (WGM) microresonators \cite{matsko09prl} where SAWs form high-quality (-Q) mechanical WGMs \cite{shui88us}. Feasibility of such interaction was proven experimentally using fused silica microresonators \cite{zehnpfennig10fio}. In this Letter we report on an experimental observation of opto-mechanical oscillation (OMO) \cite{kippenberg05prl} involving SAWs localized in crystalline lithium tantalate as well as magnesium fluoride WGM optical microresonators, and show that the monochromatic coherent SAW generated in the OMO process produces SAW pulses and SAW frequency combs. The process is similar to the optical Kerr frequency comb generation in monolithic microresonators \cite{kippenberg11s} and complements the experiments on realization of microresonator based phonon lasers \cite{grudinin10prl}.

Two experiments involving different WGM resonators were performed to validate different kinds of the OMO process. In the first experiment we used a Z-cut lithium tantalate (LiTaO$_3$) WGM resonator and scanned the relative frequency between its TE and TM WGMs by applying a DC voltage to electrodes on the top and bottom surfaces of the resonator. At at several well defined values of the DC voltage corresponding to specific frequencies fulfilling the resonant condition we observed an OMO in which the SAW wave was coupled to two nearly {\em orthogonally} polarized optical modes. In the second experiment we used a magnesium fluoride (MgF$_2$) WGM resonator. We tuned the relative frequency between the nearly {\em identically} polarized modes (either TE or TM) with temperature and observed much more efficient OMO than in the first experiment.

\begin{figure}
\centerline{\includegraphics[width=5cm]{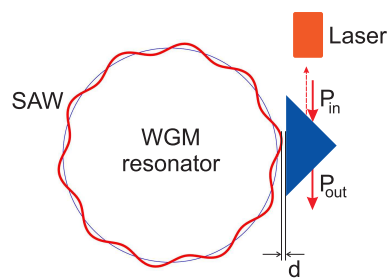}}
\caption{\label{figure7} Schematic of the experiment. }
\end{figure}

Schematic of our experiment is shown in Fig.~\ref{figure7}. A selected mode of a WGM resonator is pumped with a continuous wave light emitted by a semiconductor laser self-injection locked to the mode \cite{liang10ol}. The light is coupled in and out of the resonator using an evanescent field coupler (coupling prism). The coupling efficiency is managed by changing the distance between the coupler and the surface of the resonator, $d$. The pumping light generates a coherent acoustic wave in a SAW WGM  \cite{matsko09prl} as well as Stokes light in a properly detuned optical WGM if the power of the pump exceeds a certain threshold value.

In the first experiment we used a LiTaO$_3$ WGM resonator with $2a=0.997$~mm diameter and 0.1~mm thickness. We sent 2.3~mW of 1550~nm continuous wave light to a TE WGM mode (5~MHz loaded full width at half maximum, FWHM, $Q=4\times10^7$) and observed generation of Stokes sidebands in a TM WGM (4.3~MHz FWHM).

\begin{figure}
\centerline{\includegraphics[width=7cm]{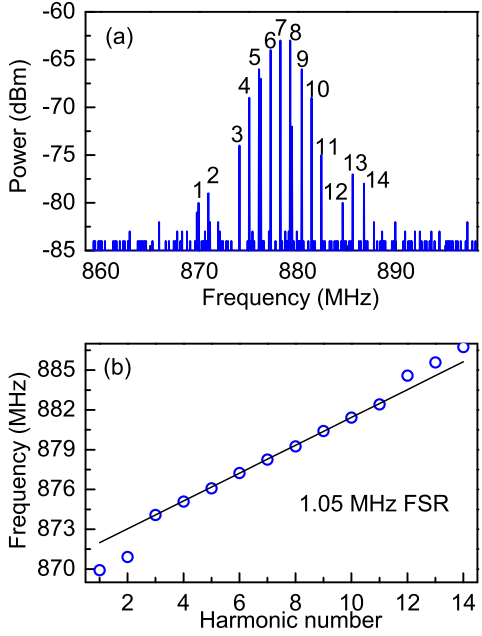}}
\caption{\label{figure1} Measurement of the FSR of the SAW WGMs
using Stokes light emitted by the resonator. The emission is observed when the frequency
difference of the TE and TM modes coincide with corresponding
harmonic of SAW WGM, and the pump power is strong enough to initialize
the OMO. The harmonics (a) are measured using the RF beat note between
the pump and Stokes light demodulated with a photodiode and recorded
with an RF spectrum analyzer. The spectrum of harmonics is
equidistant (b). Three families of lines, belonging to
three different SAW WGM families, are observed.}
\end{figure}

To quantify the observed oscillation frequencies we introduced the light escaping the resonator to a fast photodiode (50~GHz bandwidth U2t) and an RF spectrum analyzer. The resultant spectrum was centered around 880~MHz and is shown in Fig.~(\ref{figure1}). By measuring the frequency of the individual RF lines we found the free spectral range (FSR) of the acoustic modes to be $\nu_{FSR}=1.05$~MHz and the speed of the waves, given by $V_{SAW}=2\pi a \nu_{FSR}=3.3$~km$/$s, which corresponds to the tabulated speed of surface acoustic waves in LiTaO$_3$.

\begin{figure}
\centerline{\includegraphics[width=8cm]{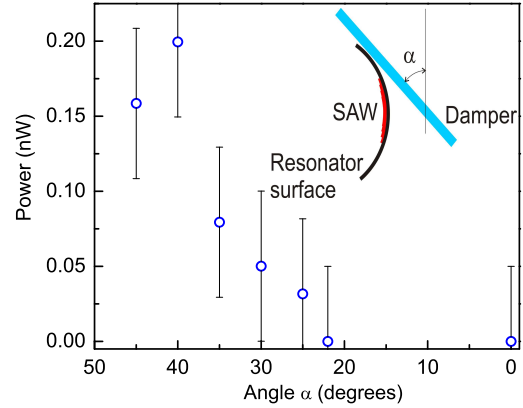}}
\caption{\label{figure2} Dependence of the RF signal produced by the
modulated light escaping the resonator on a fast photodiode, on the
position of the fused silica plate touching the resonator. The
signal disappears when the angle is less than 20 degrees. This confirms
that the SAW WGM is localized between $\alpha=\pm 20^o$ at the
resonator surface.}
\end{figure}

The SAW WGM is tightly confined by the boundaries of the resonator. We identified the confinement area by touching the surface of the resonator with a polished fused silica plate. The plate did not degrade the Q-factor of the optical modes since its index of refraction is significantly lower compared with the one of LiTaO$_3$. However, it suppressed the SAW as well as the OMO process, leading to the reduction of the power of the RF signal generated at the photodiode. We quantified the effect using the strongest line in the manifold shown in Fig.~(\ref{figure1}).  The results of the measurements, shown in Fig.~(\ref{figure2}), confirm that the SAW WGM is localized in the vicinity of the circumference of the resonator.

The OMO observed in LiTaO$_3$ resonator is not very efficient since
phase matching between the modes of the opposite polarization
used in the experiment is incomplete and the Q-factor of the resonator is not
very high. We have repeated the experiment with a resonator made out
of MgF$_2$. The resonator was shaped as a
truncated spheroid with radius $2a=2.4$~mm and thickness $0.2$~mm. In such
a structure there are no pure TE and TM modes. Modes belonging to,
say, the TE mode family have a slight deviation in the direction of
their polarization depending on the mode numbers. The crystal is
anisotropic and, hence, the modes have different thermal
sensitivity. It is possible to
reconfigure the entire spectrum of the resonator by changing its temperature. The process is less controllable compared with the electro-optical tuning of TE and
TM modes in LiTaO$_3$ resonator, yet it allows finding two
optical and a SAW WGM for the OMO.

The MgF$_2$ resonator had $Q=3\times 10^9$ (67~kHz FWHM) loaded Q-factor, which
ensured high efficiency of the interaction. The resonator was glued to the
substrate to suppress the volumetric mechanical vibrations; only SAW
WGMs survive in such a structure. We pumped the resonator with 1.7~mW of 1543~nm cw light. The laser was self-injection locked to a selected optical WGM.

\begin{figure}
\centerline{\includegraphics[width=8cm]{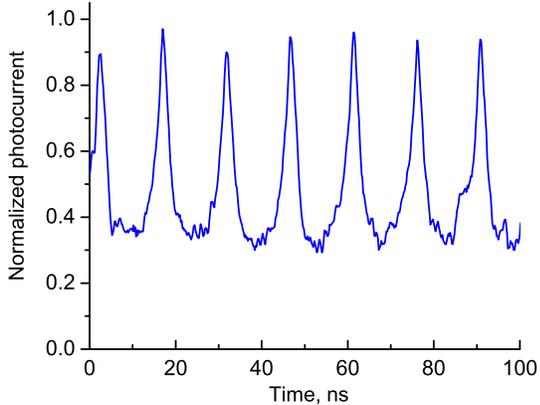}}
\caption{\label{figure3} Waveform of the optical pulses escaping the
MgF$_2$ resonator recorded with the fast oscilloscope.  }
\end{figure}

We observed generation of optical pulses shown in Fig.~\ref{figure3} by sending the light escaping the resonator to a fast photodiode directly connected to the 50~Ohm
input of a fast oscilloscope (Tektronix 300~MHz). The value of the optical power at
the input of the photodiode was 0.11~mW, which is far from
its saturation power of 20~mW. Also, the frequency of the generated RF signal
(67~MHz) was much smaller than the 50 GHz bandwidth of the photodiode.
Therefore, the measurement system operated in a linear regime and
the observed pulses were originated by the resonator, not the measurement
system. The signal disappeared if the laser was detuned from the
optical WGM or the temperature of the resonator changed by about two
degrees.

The Fourier transform of the generated pulse train reveals a number of equidistant harmonics separated by 67~MHz. The spectrum is significantly wider than the bandwidth of the optical WGMs and scales up to 1~GHz. The spectrum certainly cannot be generated within the optical WGM. Instead, it is generated due to the modulation of the coupling occurring when the SAW pulse passes by the evanescent field coupler (Figure~\ref{figure7}). The quality factor $Q_c$ describing the loading efficiency of the fundamental WGM family depends on the on the gap between the resonator surface and the coupling prism, $d$, as
\begin{equation}
Q_c\approx \frac{\pi (n_r^2-1)l^{3/2}}{2n_r \sqrt{n_c^2-n_r^2}} \exp \left [\frac {4 \pi d }{\lambda} \sqrt{n_r^2-1}\right]
\end{equation}
where $n_r$ and $n_c$ the refractive indices of the resonator and the coupling prism, and $\lambda$ and $l$ are the wavelength and the number of the mode. Therefore, modulation of the gap $d$ results in the modulation of the power of the light reflected from the prism coupler. Our observation supports the conclusion that resonant modulators with high Q-factors can operate at frequencies much larger than the resonator linewidth if the modulation of light is realized via control of the coupling efficiency \cite{sacher08oe}.

A monochromatic SAW cannot generate the observed wide spectrum, even if we take into account the nonlinear dependence of the coupling efficiency on $d$. The only possibility is that the acoustic pulses are formed at the surface of the resonator. This is a consistent picture since the SAW WGMs have equidistant spectra, and also are characterized with cubic nonlinearity \cite{mayer95pr}. A monochromatic SAW is generated due to the opto-mechanical effect, and then becomes strong enough to generate an acoustic frequency comb similarly to the generation of the optical Kerr frequency comb in a nonlinear WGM resonator pumped with cw light \cite{kippenberg11s}.

\begin{figure}
\centerline{\includegraphics[width=8cm]{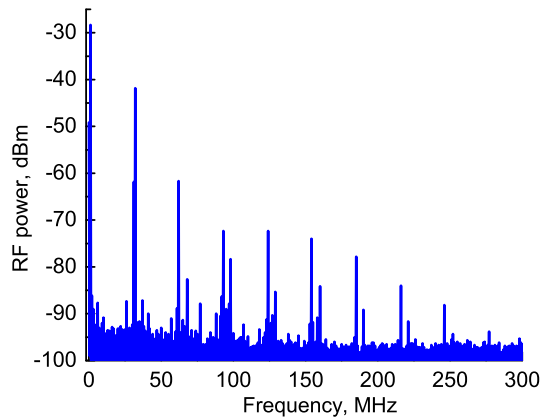}}
\caption{\label{figure5} RF spectrum generated at the fast
photodiode by the light leaving the WGM resonator. The power of the
light at the photodiode was less than 1~mW which is far from the
saturation power of the photodiode. We selected 1~MHz resolution
bandwidth in this measurement.}
\end{figure}

We performed a direct demonstration of the broad RF spectrum
by selecting another optical mode and a different temperature for the
resonator. The light exiting the optically pumped resonator was sent
to the  fast photodiode connected to a 26.5~GHz Agilent RF spectrum
analyzer. The observed RF spectrum (Figure~\ref{figure5}) was
much wider than the full width at half maximum of any
of the optical WGMs. The Q-switching mechanism explains such a broad spectrum \cite{sacher08oe}.

To conclude, we have demonstrated a triply-resonant opto-mechanical
interaction between two optical and one SAW WGMs. The process takes
place when the optical modes have either the same or orthogonal
polarizations. The large magnitude of the surface acoustic wave excited in the
OMO results in nonlinear generation of  SAW frequency
combs, leading to a Q-switching phenomenon at the output coupling port of the resonator, and in the generation of much shorter optical pulses than allowed by the intrinsic quality factor
of the optical modes. The observed phenomena are promising for
creation of extremely efficient novel opto-mechanical systems for the study of, for example, efficient optical cooling of acoustic modes down to the zero phonon level.

\end{document}